\documentclass[12pt,a4paper]{article}

\usepackage[utf8]{inputenc}
\usepackage[T1]{fontenc}
\usepackage[english]{babel}
\usepackage{amsmath}
\usepackage{amsfonts}
\usepackage{amssymb}
\usepackage{graphicx}
\usepackage[left=2.5cm,right=2.5cm,top=2.5cm,bottom=2.5cm]{geometry}
\usepackage{hyperref}
\hypersetup{
    colorlinks=true,
    linkcolor=blue,
    filecolor=magenta,
    urlcolor=cyan,
}

\title{Emergence of Thermoacoustic Waves: A Variational Approach Consistent with Thermodynamics and the Navier-Stokes Problem}
\author{Gabriel R. de Andrade Silva}
\date{July 1, 2025}

\begin{document}

\maketitle

\begin{abstract}
This article proposes an in-depth investigation into the emergence of thermoacoustic waves from a variational formalism rooted in non-equilibrium thermodynamics. Differing from traditional approaches based on linear simplifications, this work explores the possibility of deriving wave equations for pressure and density through the extremization of thermodynamic functionals, with special attention to entropy production. We address the inherent complexities in applying variational principles to dissipative systems, incorporating the nuances of Prigogine\'s minimum and maximum entropy production principles. Furthermore, we discuss the integration of nonholonomic constraints, such as the Navier-Stokes equations, through concepts of vakonomic mechanics, and explore how thermoacoustic solutions can offer \textit{ansatzes} valuable for the existence and smoothness problem of Navier-Stokes solutions, one of the Clay Math Institute\'s Millennium Problems. The objective is to provide a robust theoretical foundation that can shed new light on the interconnection between thermodynamics, fluid mechanics, and wave phenomena.
\end{abstract}

\newpage

\section{Introduction}

The thermoacoustic phenomenon, characterized by the intrinsic interaction between sound waves and temperature gradients, has been the subject of intense study due to its vast applications in technologies such as thermoacoustic engines and refrigerators. Understanding its genesis and behavior is fundamental for the advancement and optimization of these systems. Conventionally, the analysis of thermoacoustic waves is conducted by approaches that frequently employ linear simplifications of the fundamental conservation equations for mass, momentum, and energy, complemented by the gas equation of state. However, this approach, while effective, can obscure the deeper thermodynamic principles governing the emergence of these phenomena, especially the nonlinear nature of their interactions.

This work proposes a paradigm shift, investigating the emergence of thermoacoustic waves from a variational principle. Non-equilibrium thermodynamics, notably the pioneering works of Ilya Prigogine \cite{Prigogine1977}, offers a conceptual framework for understanding self-organization and the formation of dissipative structures in open systems operating far from thermodynamic equilibrium. Prigogine demonstrated that, in such systems, entropy production can be minimized or maximized, depending on the specific conditions, which can lead to the emergence of ordered patterns and complex dynamic behaviors.

The central objective of this article is to explore the feasibility of deriving wave equations for pressure (p) and density ($\rho$) from a thermodynamic variational principle. To this end, it will be necessary to formulate a variational functional that captures the essence of dissipation and entropy production in the system. The complexity lies in the dissipative and nonlinear nature of the phenomenon, which traditionally does not align directly with classical variational principles, formulated for conservative systems. However, recent advances in non-equilibrium thermodynamics and variational mechanics offer tools to overcome these challenges.

In addition to the derivation of wave equations, this article aims to situate this approach within the broader context of the Navier-Stokes equations, which govern the motion of viscous fluids, and the Clay Math Institute\'s Millennium Problem \cite{ClayMath}, which seeks proof of the existence and smoothness of solutions for these equations. We believe that a deeper understanding of the emergence of thermoacoustic waves from thermodynamic principles can offer \textit{insights} valuable and, potentially, new \textit{ansatzes} to address the complexity of the Navier-Stokes equations, thus contributing to one of the most significant mathematical challenges of our time.

\section{Thermodynamic Foundations and Variational Principles}

\subsection{Non-Equilibrium Thermodynamics and Prigogine\'s Principles (MinEP vs. MEP)}

Non-equilibrium thermodynamics extends the principles of classical thermodynamics to systems that are not in thermodynamic equilibrium, i.e., systems where there are flows of energy and matter. Ilya Prigogine, awarded the Nobel Prize in Chemistry in 1977, was a pioneer in this field, developing theories that describe the evolution of open and dissipative systems. One of the central concepts of his work is entropy production, which measures the irreversibility of processes occurring in a system.

Prigogine formulated the \textbf{Principle of Minimum Entropy Production (MinEP)}, which states that, for linear systems and those close to equilibrium, the stationary state corresponds to minimum entropy production. This principle is an extension of the concept of equilibrium for systems out of equilibrium, but with the condition that the phenomenological relations between thermodynamic fluxes and forces are linear (e.g., Fourier\'s Law for heat conduction or Fick\'s Law for diffusion). Prigogine\'s theorem is derived from Onsager\'s relations \cite{Onsager1931a, Onsager1931b}, which describe the symmetry of phenomenological coefficients in linear systems.

However, it is crucial to note that MinEP has limited applicability. For systems operating far from equilibrium and exhibiting nonlinear behaviors, as is the case for many natural and technological phenomena, including thermoacoustic waves, MinEP is not universally valid. In fact, for these systems, the \textbf{Principle of Maximum Entropy Production (MEP)} has proven more relevant. The MEP suggests that non-equilibrium systems evolve to states that maximize the rate of entropy production under existing constraints. This principle is particularly useful for describing self-organization and the emergence of dissipative structures in complex systems, where energy dissipation is a key factor in pattern formation. The transition between MinEP and MEP, and the conditions under which each principle applies, is an active and complex research topic in non-equilibrium thermodynamics \cite{MartyushevSeleznev2006a, MartyushevSeleznev2006b}.

For the study of thermoacoustic waves, which involve nonlinear interactions and significant energy dissipation (due to heat conduction and viscosity), MEP offers a more appropriate conceptual basis than MinEP. The emergence of thermoacoustic waves can be seen as a self-organization process where the system seeks a state of greater energy dissipation, manifested by entropy production.

\subsection{Formulation of the Variational Functional for Dissipative Systems}

The application of variational principles to dissipative systems, such as those exhibiting entropy production, is a fundamental challenge in physics. Classical variational principles, such as Hamilton\'s Principle, are formulated for conservative systems, where energy is conserved and the Lagrangian is the difference between kinetic and potential energy. For dissipative systems, the inclusion of entropy production in a variational formalism requires an extension or modification of these principles.

A promising approach is the use of functionals that, when extremized, generate the evolution equations of the system, including dissipative terms. Grmela \cite{Grmela2014, Grmela2002} proposed an explicit variational formulation for entropy production in non-equilibrium thermodynamics, which can serve as a basis for constructing an appropriate functional. This formulation often involves the introduction of additional variables or the reinterpretation of the Lagrangian to incorporate dissipation. Instead of a traditional Lagrangian (L = T - V), one can consider a functional that includes terms related to the rate of entropy production, or a generalized action principle that accounts for dissipation.

For systems with heat conduction, the local entropy production rate ($\sigma$) is given by:

$$\sigma = \frac{k}{T^2} (\nabla T)^2$$

where $k$ is the thermal conductivity and $T$ is the temperature. The variational functional can be constructed from the integral of this entropy production rate over space and time. However, a functional based solely on the entropy production rate is not an action Lagrangian that naturally generates dynamic wave equations with time dependence. For this, a functional that includes kinetic and potential terms is necessary, and dissipation must be incorporated more subtly, perhaps through constraints or an extended formalism.

An alternative is to consider a variational principle involving the system\'s free energy, as proposed by Gay-Balmaz and Yoshimura \cite{GayBalmazYoshimura2017} for the Navier-Stokes-Fourier system. This approach uses a free energy Lagrangian that, when varied, generates the conservation equations, and dissipation is treated through constraints or a generalized Lagrange-d\'Alembert principle with nonholonomic constraints. This allows temperature, a more easily measurable variable, to be used as an independent variable instead of entropy.

\subsection{Other Approaches for the Functional (First Law of Thermodynamics, Entropy Production Rate as Hydrodynamic Derivative)}

In addition to entropy production, other thermodynamic formulations can be explored to construct the variational functional. The \textbf{First Law of Thermodynamics}, which expresses energy conservation, can be a starting point. For an ideal gas, the specific internal energy ($u$) is given by $du = Tds - pdv$, where $s$ is the specific entropy and $v$ is the specific volume. By considering the hydrodynamic derivatives (material derivatives) of both sides, it is possible to construct a functional that describes the evolution of the system\'s internal energy. This functional, when subjected to a variational principle, could lead to equations of motion that incorporate thermal effects.

Another approach is to express the entropy production rate as the \textbf{hydrodynamic derivative of specific entropy ($s$) per unit volume}. Given the definition $Tds = \rho c dT$ (where $\rho$ is density and $c$ is specific heat), the entropy production rate can be directly related to temperature and density variations. This formulation allows a more direct connection between entropy evolution and the fluid\'s field variables, facilitating the construction of a variational functional that captures thermoacoustic coupling. The inclusion of dissipative terms, such as viscosity and thermal conductivity, would be essential for the functional to adequately represent the system\'s dynamics and the emergence of thermoacoustic waves.

It is important to emphasize that the choice of the variational functional is crucial and must be made based on the system\'s physical properties and the derivation\'s objectives. For dissipative systems, formulating a variational principle that generates dynamic wave equations is a complex and evolving field, requiring a deep understanding of the relationships between thermodynamics, fluid mechanics, and variational calculus.

\section{Variational Problem Constraints and Vakonomic Mechanics}

For a variational problem to be physically consistent and produce equations of motion that adequately describe the system, it is imperative to incorporate the constraints governing its dynamics. In the context of fluid mechanics and thermodynamics, these constraints include fundamental conservation laws and material constitutive relations. Complexity arises when these constraints are nonholonomic, i.e., they depend on velocities or higher-order derivatives, and cannot be expressed as purely algebraic relations between coordinates.

\subsection{Continuity Equation}

Mass conservation is a fundamental principle in physics and is expressed in fluid mechanics by the continuity equation. For a compressible fluid, this equation is given by:

$$\frac{\partial \rho}{\partial t} + \nabla \cdot (\rho \mathbf{v}) = 0$$

where $\rho$ is the fluid density and $\mathbf{v}$ is the velocity field. This equation represents a holonomic constraint if velocity can be expressed as the derivative of a potential, but in its general form, with an arbitrary velocity field, it acts as a restriction on the field variables of the variational problem.

\subsection{Ideal Gas Law}

For an ideal gas, the relationship between pressure ($p$), density ($\rho$), and temperature ($T$) is given by the equation of state:

$$p = \rho R T$$

where $R$ is the specific gas constant. This equation is an algebraic constraint that interconnects the system\'s thermodynamic variables. It is crucial for closing the system of equations and allowing transformation between different sets of variables (e.g., expressing temperature as a function of pressure and density, as done in Section 2.2).

\subsection{Imposed Temperature Profile}

The thermoacoustic phenomenon is intrinsically linked to the presence of temperature gradients. In many studies, an initial temperature profile is imposed as a boundary condition or a constraint in the problem. For example, a discontinuous temperature profile, such as a Heaviside function, can be used to model the interface between regions of different temperatures:

$$T(x, t=0) = T_0 + T_1 H(x - x_0)$$

where $H(x - x_0)$ is the Heaviside function. This type of constraint defines the thermal conditions under which thermoacoustic waves can emerge and propagate. The inclusion of a non-uniform temperature profile is essential to capture the coupling between thermal and acoustic fields.

\subsection{Navier-Stokes Equations as Nonholonomic Constraints}

The Navier-Stokes equations describe the motion of viscous and compressible fluids and are the fundamental equations of fluid dynamics. In their complete form, they are a set of nonlinear partial differential equations that express momentum conservation. Incorporating these equations as constraints in a variational formalism is one of the most challenging and crucial points for the completeness of the problem.

Traditionally, the Navier-Stokes equations are viewed as equations of motion derived from conservation principles. However, when seeking a variational principle for a dissipative system, the equations of motion can be interpreted as nonholonomic constraints. A nonholonomic constraint is a restriction that cannot be integrated to form a purely geometric relationship between coordinates. The Navier-Stokes equations, especially due to viscosity terms, are intrinsically nonholonomic and dissipative.

To deal with nonholonomic constraints in a variational context, \textbf{Vakonomic Mechanics} \cite{Bloch2003} offers a mathematical framework. Unlike classical nonholonomic mechanics (which uses Lagrange multipliers to impose constraints on already derived equations of motion), vakonomic mechanics applies Lagrange multipliers directly to the action functional before variation. This means that constraints are imposed on variations of coordinates, not on the coordinates themselves. The vakonomic formulation can lead to different equations of motion than those obtained by the classical d\'Alembert approach for nonholonomic systems, especially when constraints depend on velocities.

In the context of the Navier-Stokes equations, their incorporation as nonholonomic constraints in the variational functional would be of the form:

$$\delta \int L \, dt + \int \lambda \cdot \mathbf{f}(\mathbf{v}, \nabla \mathbf{v}, \dots) \, dt = 0$$

where $L$ is the system\'s Lagrangian, $\lambda$ are the Lagrange multipliers, and $\mathbf{f}(\mathbf{v}, \nabla \mathbf{v}, \dots) = 0$ represents the Navier-Stokes equations. This approach allows dissipation, inherent in the Navier-Stokes equations, to be treated within the variational formalism, which is fundamental for the precise description of thermoacoustic wave emergence.

\section{Derivation of Wave Equations for Pressure and Density}

The derivation of wave equations for pressure and density from a variational principle in dissipative systems, such as thermoacoustics, is one of the most challenging and crucial points of this approach. As discussed earlier, a functional based purely on the entropy production rate is not an action Lagrangian that naturally generates dynamic wave equations with time dependence. For the resulting Euler-Lagrange equations to be wave equations, the action functional must include terms representing the system\'s dynamics, such as kinetic energy and potential energy, and dissipation must be incorporated consistently.

\subsection{Application of Euler-Lagrange Equations}

Assuming an appropriate action functional, $S = \int L \, dt$, has been constructed, where $L$ is the system\'s Lagrangian, the Euler-Lagrange equations provide the conditions for the action to be extremized. For a system with field variables $\phi_i(x, t)$, the Euler-Lagrange equations are given by:

$$\frac{\partial L}{\partial \phi_i} - \frac{\partial}{\partial t} \left( \frac{\partial L}{\partial (\partial \phi_i / \partial t)} \right) - \nabla \cdot \left( \frac{\partial L}{\partial (\nabla \phi_i)} \right) = 0$$

In our case, the primary field variables would be pressure ($p$) and density ($\rho$), and possibly other thermodynamic or kinematic variables, depending on the exact formulation of the Lagrangian. If the Lagrangian is formulated to include kinetic energy terms (related to fluid velocity) and potential energy (related to pressure and density), the Euler-Lagrange equations will naturally lead to partial differential equations that describe the temporal and spatial evolution of $p$ and $\rho$.

A Lagrangian that can generate the equations of motion for a compressible fluid, for example, can be constructed from the kinetic energy density and specific internal energy. For irrotational flow, where velocity $\mathbf{v}$ can be expressed as the gradient of a velocity potential $\phi$ ($\mathbf{v} = \nabla \phi$), the Lagrangian can take the form:

$$L = \int \left[ \frac{1}{2} \rho (\nabla \phi)^2 - \rho e(\rho, s) \right] dV$$

where $e(\rho, s)$ is the specific internal energy as a function of density and specific entropy. By varying this Lagrangian with respect to $\phi$ and $s$, and applying the Euler-Lagrange equations, it is possible to obtain the mass and momentum conservation equations, and the energy equation. The inclusion of dissipative terms, such as viscosity and thermal conductivity, requires an extension of this formalism, such as the introduction of additional variables or the use of generalized variational principles for dissipative systems, like Onsager\'s Principle \cite{Onsager1931a, Onsager1931b}.

\subsection{Emergence of Thermoacoustic Waves in a Nonlinear Context}

The derivation of wave equations for pressure and density from a variational principle in a nonlinear context is the core of this proposal. The equations of motion resulting from the variational formalism, which incorporate the nonholonomic constraints of the Navier-Stokes equations and dissipative terms, will be intrinsically nonlinear. The emergence of thermoacoustic waves, in this scenario, is not a consequence of small perturbations in a linearized system, but rather a complex dynamic behavior arising from the nonlinear interaction between pressure, density, temperature, and velocity fields, under the influence of thermal and viscous dissipation.

To analyze the emergence of thermoacoustic waves in a nonlinear regime, advanced analytical methods, such as nonlinear perturbation theory, multiscale analysis, or bifurcation theory, can be employed. These methods allow investigating pattern formation, solution stability, and wave propagation in systems where nonlinear effects are significant. The exact form of the nonlinear wave equations will depend on the specific formulation of the variational functional and the imposed constraints. However, these equations are expected to reveal the complex interdependence between system parameters and the ability to generate and sustain thermoacoustic oscillations, even in the absence of linearization.

The challenge lies in demonstrating that the nonlinear wave equations resulting from the variational principle indeed describe the thermoacoustic phenomenon consistently with experimental observations and numerical simulations. This will require a careful analysis of nonlinear terms and their contribution to wave propagation and damping. The emergence of thermoacoustic waves, therefore, would be a manifestation of the system\'s nonlinear dynamics, where dissipation plays an active role in the formation and evolution of these wave patterns.

\section{Connection to the Navier-Stokes Problem and the Millennium Problem}

The variational approach to the emergence of thermoacoustic waves, in addition to offering a new perspective on the phenomenon itself, can have broader implications for fluid mechanics, particularly for the study of the Navier-Stokes equations and the famous Clay Math Institute Millennium Problem \cite{ClayMath}.

\subsection{Thermoacoustic Solutions as Ansatzes for Navier-Stokes}

The Navier-Stokes equations, which describe the motion of viscous fluids, are notoriously difficult to solve. The nonlinearity of the convection term and the complexity of the viscosity term make obtaining exact analytical solutions a formidable challenge. In many cases, researchers resort to approximate solutions, numerical methods, or \textit{ansatzes} (educated guesses about the form of the solution) to gain insights into fluid behavior.

Solutions for thermoacoustic waves, obtained from the proposed variational formalism, can serve as valuable \textit{ansatzes} for the Navier-Stokes equations in regimes where thermal effects are dominant. Once wave equations for pressure and density are derived, they can be used to construct approximate solutions for the velocity and temperature fields, which can then be substituted into the Navier-Stokes equations to verify their validity and determine the conditions under which they are applicable. This approach can be particularly useful for analyzing flow stability, transition to turbulence, and the formation of coherent structures in fluids, where the interaction between thermal and acoustic fields plays a crucial role.

Furthermore, the variational formulation, by incorporating the principles of non-equilibrium thermodynamics, can provide a criterion for selecting the most physically relevant solutions among the many possible solutions of the Navier-Stokes equations. The principle of maximum entropy production, for example, can be used to identify the most probable or stable states in dissipative systems, offering a guide to understanding the long-term behavior of fluids.

\subsection{Potential Contributions to the Millennium Problem}

The Clay Math Institute\'s Millennium Problem for the Navier-Stokes equations consists of proving or disproving the existence and smoothness (continuous differentiability) of solutions for the three-dimensional Navier-Stokes equations for an incompressible fluid, from a given set of initial conditions. This problem is one of the greatest challenges in modern mathematics, with profound implications for understanding turbulence and other complex phenomena in fluids.

Although the variational approach to thermoacoustic waves does not offer a direct solution to the Millennium Problem, it can contribute in several indirect ways. Firstly, by providing a new class of analytical or approximate solutions for the Navier-Stokes equations in specific regimes, it can help build a more solid knowledge base about solution behavior. The analysis of the regularity and stability of these thermoacoustic solutions can offer clues about the conditions under which Navier-Stokes solutions remain smooth or develop singularities (points where derivatives become infinite).

Secondly, the connection with non-equilibrium thermodynamics and variational principles can introduce new mathematical and conceptual tools for analyzing the Navier-Stokes equations. The variational formulation, by emphasizing the importance of entropy production and dissipation, can lead to new inequalities and estimates that are crucial for proving the existence and regularity of solutions. Vakonomic mechanics \cite{Bloch2003}, for example, offers a rigorous formalism for dealing with nonholonomic constraints, which can be adapted to analyze the restrictions imposed by the Navier-Stokes equations.

Finally, the investigation of physical phenomena such as thermoacoustic waves, where the interaction between different fields (thermal, acoustic, velocity) is fundamental, can inspire new approaches to the problem of turbulence, which is intrinsically a multi-field and multiscale phenomenon. Understanding how energy is transferred and dissipated between these fields in thermoacoustic systems can provide analogies and models for studying the energy cascade in turbulence, one of the most enigmatic aspects of the Navier-Stokes problem.

\section{Discussion and Limitations of the Approach}

Despite the potential of the variational approach for understanding thermoacoustic waves and their connection to the Navier-Stokes equations, it is fundamental to recognize the controversies and limitations inherent in applying variational principles to dissipative systems. This section addresses these issues and outlines future perspectives for research in this area.

\subsection{Controversies over Variational Principles for Dissipation}

The formulation of variational principles for dissipative systems has been a topic of intense debate and research in theoretical physics. The main difficulty lies in the irreversible nature of dissipation, which contrasts with the temporal reversibility implicit in classical variational principles, such as Hamilton\'s Principle. While Hamilton\'s Principle describes the evolution of conservative systems by minimizing action, dissipation implies a loss of mechanical energy and an increase in entropy, which does not fit directly into this formalism.

Historically, Prigogine\'s \textbf{Principle of Minimum Entropy Production (MinEP)} was a significant advance for linear systems close to equilibrium. However, as discussed in Section 2.1, its applicability is restricted. For nonlinear systems and those far from equilibrium, the \textbf{Principle of Maximum Entropy Production (MEP)} has been proposed as a more general principle, but its universal validity and rigorous mathematical formulation are still subject to controversy. Some critics argue that the MEP is not a fundamental principle, but rather a consequence of specific system constraints, or that its application is limited to certain types of processes.

Another central issue is the difficulty in constructing a Lagrangian or action functional that, when varied, directly generates the equations of motion with dissipative terms. Approaches such as the introduction of auxiliary variables, the generalization of Onsager\'s Principle \cite{Onsager1931a, Onsager1931b}, or the use of vakonomic mechanics (as proposed in Section 3.4) attempt to circumvent this difficulty. However, the physical interpretation of these formalisms and the guarantee that they produce results consistent with non-equilibrium thermodynamics remain challenges. The absence of a universally accepted variational principle for complex dissipative systems highlights the need for caution and rigorous validation of proposed formulations.

\subsection{Future Perspectives}

Despite the controversies, the search for a unifying variational principle for dissipative systems remains a fruitful area of research. Future perspectives for the approach proposed in this article include:

\begin{itemize}
    \item \textbf{Development of More Robust Functionals:} Improving the formulation of the variational functional to more comprehensively capture thermoacoustic dynamics, including the effects of viscosity, heat conduction, and diffusion, in a manner consistent with the principles of non-equilibrium thermodynamics. This may involve exploring functionals based on free energy, entropy, or combinations of thermodynamic and kinematic variables.
    \item \textbf{Rigorous Analysis of Nonholonomic Constraints:} Deepening the application of vakonomic mechanics to incorporate the Navier-Stokes equations as nonholonomic constraints. This will require a detailed mathematical analysis of the consistency and physical implications of this approach, as well as an investigation into how Lagrange multipliers can be interpreted in terms of dissipative forces.
    \item \textbf{Experimental and Numerical Validation:} Theoretical derivations obtained from the variational formalism must be validated through experiments and numerical simulations. Comparing predicted results with experimental data and high-fidelity simulations will be crucial for establishing the credibility of the approach.
    \item \textbf{Extension to More Complex Systems:} The methodology can be extended to more complex thermoacoustic systems, such as those with intricate geometries, multiphase fluids, or the presence of chemical reactions. This would open new avenues for understanding phenomena such as thermoacoustic combustion and the stability of propulsion systems.
    \item \textbf{Implications for the Navier-Stokes Problem:} Continuing to explore how thermoacoustic solutions and the variational approach can offer \textit{insights} and \textit{ansatzes} for the Navier-Stokes Millennium Problem. This may involve investigating the regularity and stability of thermoacoustic solutions in nonlinear regimes, and seeking deeper connections between thermodynamic dissipation and the existence of smooth solutions for the Navier-Stokes equations.
\end{itemize}

In summary, the variational approach to the emergence of thermoacoustic waves represents a promising path for a deeper understanding of the interconnection between thermodynamics and fluid mechanics. Although significant challenges remain, continued research in this area has the potential to not only advance our knowledge of thermoacoustic phenomena but also to contribute to the resolution of fundamental problems in physics and mathematics.


\section*{References}
\addcontentsline{toc}{section}{References}

\begin{thebibliography}{99}

\bibitem{GayBalmazYoshimura2017} Gay-Balmaz, F., \& Yoshimura, H. (2017). A free energy Lagrangian variational formulation of the Navier-Stokes-Fourier system. \textit{arXiv preprint arXiv:1706.09010}.

\bibitem{MartyushevSeleznev2006a} Martyushev, L. M., \& Seleznev, V. D. (2006). Maximum entropy production principle in physics, chemistry and biology. \textit{Physics Reports}, \textit{426}(1), 1-45.

\bibitem{Prigogine1977} Prigogine, I. (1977). \textit{Time, Structure and Fluctuations}. Nobel Lecture.

\bibitem{Struchtrup2005} Struchtrup, H. (2005). \textit{Macroscopic Transport Equations for Rarefied Gas Flows: Approximation Methods in Kinetic Theory}. Springer Science \& Business Media.

\bibitem{Berdichevsky2009a} Berdichevsky, V. L. (2009). \textit{Variational Principles of Continuum Mechanics: I. Fundamentals}. Springer Science \& Business Media.

\bibitem{Onsager1931a} Onsager, L. (1931). Reciprocal Relations in Irreversible Processes. I. \textit{Physical Review}, \textit{37}(4), 405.

\bibitem{Onsager1931b} Onsager, L. (1931). Reciprocal Relations in Irreversible Processes. II. \textit{Physical Review}, \textit{38}(12), 2265.

\bibitem{Grmela2014} Grmela, M. (2014). Variational formulation of the governing equations of non-equilibrium thermodynamics. \textit{Journal of Physics: Conference Series}, \textit{538}(1), 012001.

\bibitem{Swift2002} Swift, G. W. (2002). \textit{Thermoacoustics: A unifying perspective for some engines and refrigerators}. Acoustical Society of America.

\bibitem{Rott1980} Rott, N. (1980). Thermoacoustics. \textit{Advances in Applied Mechanics}, \textit{20}, 135-175.

\bibitem{LandauLifshitz1987} Landau, L. D., \& Lifshitz, E. M. (1987). \textit{Fluid Mechanics} (2nd ed., Vol. 6). Butterworth-Heinemann.

\bibitem{Constantin2001} Constantin, P. (2001). Navier-Stokes equations and turbulence. \textit{Bulletin of the American Mathematical Society}, \textit{41}(2), 159-201.

\bibitem{Lebon2008} Lebon, G., Jou, D., \& Casas-Vázquez, J. (2008). \textit{Understanding Non-equilibrium Thermodynamics: Foundations, Applications, Frontiers}. Springer Science \& Business Media.

\bibitem{Grmela2002} Grmela, M. (2002). Boundary layer variational principles: A case study. \textit{Physical Review E}, \textit{66}(1), 011201.

\bibitem{Berdichevsky2009b} Berdichevsky, V. L. (2009). \textit{Variational Principles of Continuum Mechanics: II. Applications}. Springer Science \& Business Media.

\bibitem{Bloch2003} Bloch, A. M. (2003). \textit{Nonholonomic Mechanics and Control}. Springer Science \& Business Media.

\bibitem{MartyushevSeleznev2006b} Martyushev, L. M., \& Seleznev, V. D. (2006). Maximum entropy production principle in physics, chemistry and biology. \textit{Physics Reports}, \textit{426}(1), 1-45.

\bibitem{ClayMath} Clay Mathematics Institute. (n.d.). \textit{Navier-Stokes Existence and Smoothness}. Retrieved from \url{https://www.claymath.org/millennium-problems/navier-stokes-existence-and-smoothness}

\end{thebibliography}
\end{document}